\documentclass[journal,twocolumn]{IEEEtran}   
\usepackage{graphicx}
\usepackage{epstopdf}
\usepackage{amssymb}
\usepackage{amsfonts}
\usepackage{amsmath}
\usepackage{algorithm}
\usepackage{algorithmic}
\usepackage{subeqnarray}
\usepackage{cases}
\usepackage{bm}
\usepackage{subfigure,amsmath,amssymb,cite}
\usepackage{stfloats}

\usepackage{float}

\ifCLASSINFOpdf
\else
\fi
\begin{document}

\title{Power Allocation for IRS-aided Two-way Decode-and-Forward Relay Wireless Network}

\author{Xuehui~Wang,~Peng Zhang,~Feng Shu,~Weiping Shi, and~Jiangzhou~Wang,~\emph{Fellow},~\emph{IEEE}


\thanks{This work was supported in part by the National Natural Science Foundation of China (Nos. 62071234, 62071289, and 61972093), the Hainan Major Projects (ZDKJ2021022), the Scientific Research Fund Project of Hainan University under Grant KYQD(ZR)-21008, and the National Key R\&D Program of China under Grant 2018YFB180110 \emph{(Corresponding authors: Feng Shu)}.}

\thanks{Xuehui~Wang,~Peng Zhang, and~Feng Shu are  with the School of Information and Communication Engineering, Hainan University,~Haikou,~570228, China.}

\thanks{Weiping Shi and Feng Shu are with the School of Electronic and Optical Engineering, Nanjing University of Science and Technology, 210094, China.}

\thanks{Jiangzhou Wang is with the School of Engineering, University of Kent, Canterbury CT2 7NT, U.K. Email: (e-mail: j.z.wang@kent.ac.uk).}

}

\maketitle

\begin{abstract}

 In this paper, an intelligent reflecting surface (IRS)-aided two-way decode-and-forward (DF) relay wireless network is considered, where two users exchange information via IRS and DF relay. To enhance the sum rate performance, three power allocation (PA) strategies are proposed. Firstly, a method of maximizing sum rate (Max-SR) is proposed to jointly optimize the PA factors of user $U_1$, user $U_2$ and relay station (RS). To further improve the sum rate performance, two high-performance schemes, namely maximizing minimum sum rate (Max-Min-SR) and maximizing sum rate with rate constraint (Max-SR-RC), are presented.
Simulation results show that the proposed three methods outperform the equal power allocation (EPA) method in terms of sum rate performance. In particular, the highest performance gain achieved by Max-SR-RC method is up to 45.2\% over EPA. Furthermore, it is verified that the total power and random shadow variable $X_\sigma$ have a substantial impact on the sum rate performance.

\end{abstract}

\begin{IEEEkeywords}

Intelligent reflecting surface, two-way decode-and-forward relay, power allocation, sum rate.

\end{IEEEkeywords}

\section{Introduction}

There are many thorny problems in communication networks, such as the propagation loss and multi path fading, which seriously deteriorate the communication performance \cite{2022Cx}. With the ability to intelligently adjust the propagation environment, helpful multi path can be created by intelligent reflecting surface (IRS) for coverage extension. IRS has become an emerging technology with competitive advantages over the existing technologies \cite{2020Wqq}. IRS consists of cost-effective, low-power, and passive reflecting units, which reflect the signal independently to achieve passive beamforming for signal enhancement, spectral and energy efficiency improvement \cite{2020Twk, 2022Zfh}. IRS has been widely applied to different application scenarios, e.g., physical layer security \cite{2021Ljy}, simultaneous wireless information and power transfer (SWIPT) \cite{2020Swp}, \cite{2020Pch}, mobile edge computing  \cite{2021Tb}, and covert communication \cite{2022Zxb}.

Due to the fact that IRS reflects signal with low-power consumption, which can be regarded as a passive relay. While the conventional relay is an active device with strong signal processing capability to amplify and forward (AF), decode and forward (DF) signal. Its high hardware cost and high-power consumption are inconsistent with our low-energy-consumption communication demand.

How to make full use of the advantages of IRS and relay to serve the wireless communication network is crucial, which has already attracted extensive attention from both academia and industry \cite{ 2022Wxh , 2021Ty , 2021Hc , 2021Wj }.
In \cite{2022Wxh}, an IRS-aided DF relay network was considered. To optimize the beamforming at relay station (RS) and phase at IRS, three methods of maximizing receive power, i.e., using alternately iterative structure, null-space projection based plus maximum ratio combining (MRC) and IRS element selection based plus MRC, were respectively proposed for rate improvement and coverage extension.
For normal communication between end users in \cite{2021Ty}, a two-way AF relay network assisted by two IRSs was proposed, where alternatively optimizing phase at the two IRSs to maximize sum rate. It demonstrated that the efficiency and sum rate performance surpass that of only relay-aided system.
\cite{2021Hc} presented cooperative relay networks with IRS, where reflection amplitudes changes with the discrete phase. A deep reinforcement learning (DRL) scheme with low complexity was proposed to optimize relay selection and IRS reflection coefficient. It was shown that DRL outperformed random relay selection.
In \cite{2021Wj}, the authors investigated an IRS-aided two-way AF relay network, where phase was solved by signal-to-noise ratio (SNR)-upper-bound-maximization or genetic-SNR-maximization and beamforming was achieved by optimizing beamforming or maximum-ratio beamforming. It demonstrated the proposed algorithms evidently performed better than random phase.

However, all the above literature focused on the optimization of  beamforming at RS and phase at IRS, but did not consider power allocation (PA) between users and RS. PA is an efficient way to improve the sum rate, which plays an important role in sum rate. As far as we know, there is little research work on the PA of IRS-aided two-way DF relay network. This motivates us to pay attention to the PA under the total power constraint. Our main contributions are summarized as follows:

\begin{enumerate}

\item

To improve the sum rate performance of an IRS-aided two-way DF relay network, a PA strategy of maximizing sum rate (Max-SR) method is proposed to optimize the PA factors of two users and RS. Since the constraints are non-convex, the upper bound of the product of two PA factors and the first-order Taylor approximation of quadratic of PA factor are derived, so that the constraints can be converted into convex. The simulation results prove that the proposed Max-SR method can harvest up to 13.8\% sum rate gain over equal power allocation (EPA) method.

\item

To achieve a higher sum rate, two high-performance schemes, namely maximizing minimum sum rate (Max-Min-SR) and maximizing sum rate with rate constraint (Max-SR-RC), are presented. In Max-Min-SR method, all the constraints are convex with few converted operations. In Max-SR-RC method, the first-order Taylor approximation is also applied to constraint transformation. Simulation results show that the best method among the proposed three methods and EPA is Max-SR-RC, which achieves at least 10\% rate gain. Followed by Max-Min-SR method, the sum rate gain harvested is up to 15.4\%.

\end{enumerate}

The remainder of this paper is organized as follows. In Section II, an IRS-aided two-way DF relay network is described. In Section III, three high-performance schemes for a better sum rate performance of the proposed network is demonstrated. Simulation results are presented in Section IV, and conclusions are drawn in Section V.

\emph{Notation}: Scalars, vectors and matrices are respectively represented by letters of lower case, bold lower case, and bold upper case. $(\cdot)^T$ and $(\cdot)^H$ stand for transpose and conjugate transpose, respectively. $\mathbb{E}\{\cdot\}$ and $\|\cdot\|$ denote expectation operation and 2-norm, respectively. The sign $\textbf{I}_{M}$ is the $M\times M$ identity matrix, and $\oplus$ represents exclusive or operation of two decoded symbols.

\section{System Model}
\begin{figure}[h]
\centering
\includegraphics[width=0.49\textwidth,height=0.25\textheight]{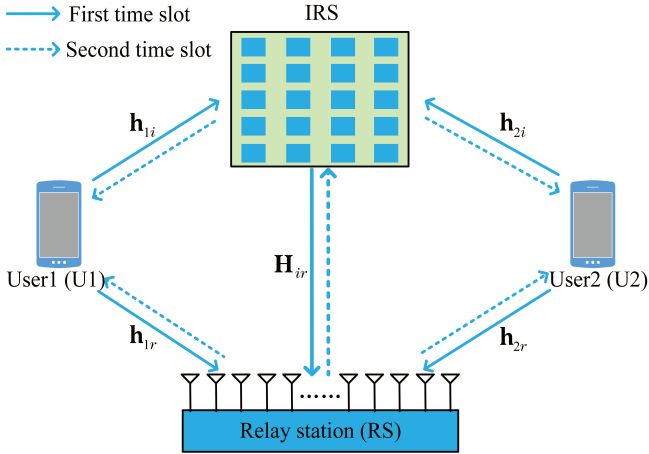}\\
\caption{System Model.}\label{System_Model.eps}
\end{figure}

As shown in Fig. 1, we consider an IRS-aided two-way relay network. The network comprises a half-duplex two-way DF RS with $M$ transmit antennas, an IRS with $N$ reflecting elements, and two single-antenna users. Two users are respectively denoted by $U_1$ and $U_2$, which mutually exchange information with the help of IRS and RS. Due to path loss, the power of signals reflected by the IRS twice or more are such weak that they can be ignored. In the first time slot, the received signal at RS can be expressed as
\begin{align}\label{y_r1}
&\textbf{y}_r=\sqrt {\beta_1{P}}( \textbf{h}_{1r}+\textbf{H}_{ir}{\boldsymbol\Theta_1}\textbf{h}_{1i})x_1  \nonumber\\
&~~~~~~~~~~~~~~~~~~~~~~+ \sqrt{\beta_2{P}}( \textbf{h}_{2r}+\textbf{H}_{ir}{\boldsymbol\Theta_1}\textbf{h}_{2i})x_2+\textbf{n}_r,
\end{align}
where $x_1$ and $x_2$ are the independent transmit signal from $U_1$ and $U_2$, respectively. $P$ is the total transmission power and limited, $\beta_1$ and $\beta_2$ are the power allocation parameters of $U_1$ and $U_2$. $\mathbb{E}\{x_1^H{x_1}\}=1$ and $\mathbb{E}\{x_2^H{x_2}\}=1$. Without loss of generality, we assume a Rayleigh fading environment \cite{20009Wjz}. Let $\textbf{h}_{1r}$$\in \mathbb C^{M \times 1}$, $\textbf{H}_{ir}\in \mathbb C^{M \times N}$ and $\textbf{h}_{1i}\in \mathbb C^{N \times 1}$ represent the channels from $U_1$ to RS, from IRS to RS, and from $U_1$ to IRS, $\textbf{h}_{2r}$$\in \mathbb C^{M \times 1}$ and $\textbf{h}_{2i}\in \mathbb C^{N \times 1}$ represent the channels from $U_2$ to RS and from $U_2$ to IRS. $\boldsymbol\Theta_1$ is the diagonal reflection-coefficient matrix of IRS, which is denoted as $\boldsymbol\Theta_1 = \text{diag}( e^{j\theta _{1}}, \cdots,e^{j\theta _{N}} )$, $\theta _{i}\in ( 0 ,{2\pi } ]$ is the phase shift of the $i$th element. $\textbf{n}_r\in \mathbb C^{M \times 1}$ is the additive white Gaussian noise (AWGN) with distribution $\textbf{n}_r\sim \mathcal{CN}( 0,\sigma_r^2{\bf I}_{M} )$.
Considering $x_2$  as unknown interference, RS first decodes $x_1$ to $\widetilde{x}_1$. Then the contribution of $x_1$ can be eliminated from (\ref{y_r1}), and $x_2$ can be decoded to $\widetilde{x}_2$.

In the second time slot, network coding is employed for $\widetilde{x}_1$ and $\widetilde{x}_2$, and combining them into a new signal, which is represented as
\begin{equation}
{x}_r=\widetilde{x}_1\oplus\widetilde{x}_2.
\end{equation}
Each user decodes ${x}_r$ to $\widetilde{x}_r$, and rebuild the symbol sent by the other user (i.e, $\overline{x}_2 = \widetilde{x}_r\oplus{x}_1$ or $\overline{x}_1 = \widetilde{x}_r\oplus{x}_2 $). It is assumed that the channel reciprocity holds, i.e., the channels in the second-time slot can be represented as the conjugate transpose of the channels in the first-time slot.
After self-interference cancellation, the received signal at $U_1$ is written by
\begin{equation}
y_1=\sqrt {\beta_3{P}}( \textbf{h}_{1r}^H+\textbf{h}_{1i}^H{\boldsymbol\Theta_2}\textbf{H}_{ir}^H)\overline{\textbf{x}}_2+n_1,
\end{equation}
where $\overline{\textbf{x}}_2 = \overline{x}_2[1, 1, \cdot\cdot\cdot,1]^T\in \mathbb C^{M \times 1}$. Similarly, the received signal at $U_2$ is given by
\begin{equation}
y_2=\sqrt {\beta_3{P}}( \textbf{h}_{2r}^H+\textbf{h}_{2i}^H{\boldsymbol\Theta_2}\textbf{H}_{ir}^H)\overline{\textbf{x}}_1+n_2,
\end{equation}
where $\overline{\textbf{x}}_1 = \overline{x}_1[1, 1, \cdot\cdot\cdot,1]^T\in \mathbb C^{M \times 1}$, $\beta_3$ is the power allocation parameter of RS, the diagonal reflection-coefficient matrix of IRS is represented as ${\boldsymbol\Theta}_2 = \text{diag}\left( e^{j{\theta _{21}}}, \cdots ,e^{j{\theta _{2N}}} \right)$, ${\theta _{2i}}\in \left( 0 \right.,\left. {2\pi } \right]$ is the phase shift of the $i$th element.
$n_1$ and $n_2$ are the AWGN with distribution $n_{1} \sim \mathcal{CN}\left( 0,{\sigma_1^2} \right)$ and $n_{2} \sim \mathcal{CN}\left( 0,{\sigma_2^2} \right)$, respectively.
The achievable rate of $U_1$-RS-$U_2$ link can be expressed as follows
\begin{equation}
R_{12}=\text{min}\{R_{1ir},R_{ri2}  \},
\end{equation}
where $R_{1ir}$ and $R_{ri2}$ are the rates of $U_1$-RS and RS-$U_2$ link, respectively, 
\begin{equation}
R_{1ir}=\frac{1}{2}\log_2 (1+   \gamma_1\beta_1{P}  ),R_{ri2}=\frac{1}{2}\log_2 (1+   \gamma_2\beta_3{P}   ),
\end{equation}
and
\begin{equation}
\gamma_1=\frac{ \| \textbf{h}_{1r}+\textbf{H}_{ir}{\boldsymbol\Theta_1}\textbf{h}_{1i} \|^2  }{\sigma_r^2},
\gamma_2=\frac{ \| \textbf{h}_{2r}^H+\textbf{h}_{2i}^H{\boldsymbol\Theta_2}\textbf{H}_{ir}^H \|^2  }{\sigma_2^2}.
\end{equation}
Similarly, the achievable rate of $U_2$-RS-$U_1$ link can be represented as follows
\begin{equation}
R_{21}=\text{min}\{R_{2ir},R_{ri1}  \},
\end{equation}
where $R_{2ir}$ and $R_{ri1}$ are the rates of $U_2$-RS and RS-$U_1$ link, respectively,
\begin{equation}
R_{2ir}=\frac{1}{2}\log_2 (1+   \gamma_3\beta_2{P}  ),
R_{ri1}=\frac{1}{2}\log_2 (1+   \gamma_4\beta_3{P}  ),
\end{equation}
and
\begin{equation}
\gamma_3=\frac{ \|  \textbf{h}_{2r}+\textbf{H}_{ir}{\boldsymbol\Theta_1}\textbf{h}_{2i} \|^2  }{\sigma_r^2},
\gamma_4=\frac{ \|  \textbf{h}_{1r}^H+\textbf{h}_{1i}^H{\boldsymbol\Theta_2}\textbf{H}_{ir}^H \|^2  }{\sigma_1^2}.
\end{equation}

The achievable multiple access channel (MAC) rate of $U_1$-RS and $U_2$-RS can be represented as follows
\begin{equation}
R_{MAC}=\frac{1}{2}\log_2 (1+   \gamma_1\beta_1{P}  +  \gamma_3\beta_2{P}).
\end{equation}
Therefore, the achievable sum rate of the proposed system is defined as follows
\begin{align}
&R=\text{min} \{  R_{12}+R_{21} , R_{MAC}  \} \nonumber\\
&=\text{min} \{ \text{min}\{R_{1ir},R_{ri2}  \}+\text{min}\{R_{2ir},R_{ri1}  \} , R_{MAC}  \}.
\end{align}

\section{Proposed three PA methods}
In this section, we focus on the investigation of PA methods to maximize $R$, and the optimization problem is casted as
\begin{subequations}
\begin{align}\label{R}
&\max \limits_{\beta_1,\beta_2,\beta_3,{\boldsymbol\Theta_1},{\boldsymbol\Theta_2}} ~~~R \\
&~~~~~~\text{s.t.}~~ 0< \beta_1, \beta_2, \beta_3 < 1,~~\beta_1+\beta_2+\beta_3=1,   \label{R_1}       \\
&~~~~~~~~~~  |\boldsymbol\Theta_1(i,i)|=1, |\boldsymbol\Theta_2(i,i)|=1, \forall i = 1, \cdots ,N,
\end{align}
\end{subequations}
where $\boldsymbol\Theta_1$ and $\boldsymbol\Theta_2$ are obtained by maximizing the sum rate via general power iterative in \cite{2022Zp}. In this paper, $\boldsymbol\Theta_1$ and $\boldsymbol\Theta_2$ refer directly to \cite{2022Zp}. The above optimization problem is simplified to
\begin{equation}
\max \limits_{\beta_1, \beta_2, \beta_3 }~~ R  ~~~~~~~~~~~~~~~~~~~~\text{s.t.}~~~~ (\text{\ref{R_1}}).
\end{equation}
To solve the optimization problem, three PA schemes are proposed, which are Max-SR, Max-Min-SR and Max-SR-RC, respectively, and the related details are as follow.

\subsection{Proposed Max-SR}
The system sum rate $R$ is expanded as follows
\begin{align}\label{R_expanded}
&R=\text{min} \{ \text{min}\{R_{1ir}+ R_{2ir}, R_{1ir}+ R_{ri1},  R_{ri2}+ R_{2ir}   \nonumber\\
&~~~~~~~~~~~~~~~~~~~~~~~~~~~~~~~~~~~, R_{ri2}+ R_{ri1} \}  , R_{MAC}  \}  \nonumber\\
&~~=\text{min} \{ R_{1ir}+ R_{2ir}, R_{1ir}+ R_{ri1},  R_{ri2}+ R_{2ir}   \nonumber\\
&~~~~~~~~~~~~~~~~~~~~~~~~~~~~~~~~~~~, R_{ri2}+ R_{ri1}  , R_{MAC}  \}.
\end{align}
Clearly, $R_{1ir}+ R_{2ir}>R_{MAC}$, so the case of $R_{1ir}+ R_{2ir}$ is excluded directly. The above equation is reduced to
\begin{equation}
R=\text{min} \{ R_{1ir}+ R_{ri1},  R_{ri2}+ R_{2ir},  R_{ri2}+ R_{ri1},  R_{MAC}  \},
\end{equation}
the optimization problem is further recasted as
\begin{subequations}
\begin{align}\label{A}
&\max \limits_{\beta_1, \beta_2, \beta_3, R }~~ R  \\
&~~~~\text{s.t.}~~~~~~~ 0< \beta_1, \beta_2, \beta_3 < 1,~~\beta_1+\beta_2+\beta_3=1,  \\
&~~~~~~~~~~~~~~  R \leq R_{1ir}+ R_{ri1},   \label{A_1}\\
&~~~~~~~~~~~~~~  R \leq R_{ri2}+ R_{2ir},    \label{A_2}\\
&~~~~~~~~~~~~~~  R \leq R_{ri2}+ R_{ri1},   \label{A_3} \\
&~~~~~~~~~~~~~~  R \leq R_{MAC},
\end{align}
\end{subequations}
where
\begin{subequations}
\begin{align}
&R_{1ir}+ R_{ri1}= \frac{1}{2}\log_2 (1+   \gamma_1\beta_1{P}  +   \gamma_4\beta_3{P} + \gamma_1\gamma_4\beta_1\beta_3{P^2}  ), \\
&R_{ri2}+ R_{2ir}= \frac{1}{2}\log_2 (1+   \gamma_3\beta_2{P}  +   \gamma_2\beta_3{P} + \gamma_2\gamma_3\beta_2\beta_3{P^2}  ), \\
&R_{ri2}+ R_{ri1}= \frac{1}{2}\log_2 [1+   \beta_3{P}(\gamma_2+\gamma_4)    + \gamma_2\gamma_4{\beta_3}^2{P^2}  ].
\end{align}
\end{subequations}
The above optimization is non-convex due to the non-convex constraints (\ref{A_1}), (\ref{A_2}) and (\ref{A_3}), it is necessary to convert (\ref{A_1}), (\ref{A_2}) and (\ref{A_3}) to convex. 
Inserting $\beta_3=1-\beta_1-\beta_2$ back into $R_{1ir}+ R_{ri1}$ yields the following inequality
\begin{align}
&2^{2R}\leq 1+\gamma_4P+\beta_1(\gamma_1P-\gamma_4P+\gamma_1\gamma_4P^2)\nonumber\\
&~~~~~~~~~~~~~~~~~~ -\gamma_1\gamma_4\beta_1^2P^2-\gamma_4\beta_2P-\gamma_1\gamma_4\beta_1\beta_2P^2.
\end{align}
Since $\beta_1\beta_2 \leq \frac{1}{2}(\beta_1^2+\beta_2^2)$, the lower-bound of $-\beta_1\beta_2$ is $-\frac{1}{2}(\beta_1^2+\beta_2^2)$, which further yields
\begin{align}\label{18d}
&2^{2R}\leq 1+\gamma_4P+\beta_1(\gamma_1P-\gamma_4P+\gamma_1\gamma_4P^2)-\gamma_4\beta_2P\nonumber\\
&~~~~~~~~~~~~~~~~~~~~~~~~~~~~~~~~~~~~~ -\frac{\gamma_1\gamma_4(3\beta_1^2+\beta_2^2)P^2}{2},
\end{align}
which is a convex constraint. In the same manner, (\ref{A_2}) can be rewritten as
\begin{align}\label{18e}
&2^{2R}\leq 1+\gamma_2P+\beta_2(\gamma_3P-\gamma_2P+\gamma_2\gamma_3P^2)-\gamma_2\beta_1P\nonumber\\
&~~~~~~~~~~~~~~~~~~~~~~~~~~~~~~~~~~~~~ -\frac{\gamma_2\gamma_3(\beta_1^2+3\beta_2^2)P^2}{2}.
\end{align}
Similarly, (\ref{A_3}) can be written in the following form
\begin{equation}\label{17f1}
2^{2R} \leq  1+\beta_3P(\gamma_2+\gamma_4)+\gamma_2\gamma_4\beta_3^2{P^2},
\end{equation}
which is still a non-convex constraint. $\gamma_2\gamma_4\beta_3^2{P^2}$ is a convex function, its low bound can be expressed by the first-order Taylor expansion. The first-order Taylor expansion of $\gamma_2\gamma_4\beta_3^2{P^2}$ at feasible point $\beta_{3t}$ is given by
\begin{equation}\label{17f2}
\gamma_2\gamma_4\beta_3^2{P^2} \geq  \gamma_2\gamma_4\beta_{3t}^2{P^2} +  2\gamma_2\gamma_4\beta_{3t}{P^2} (\beta_{3}-\beta_{3t}).
\end{equation}
Combining (\ref{17f1}) and (\ref{17f2}), (\ref{A_3}) further can be converted into
\begin{equation}\label{18f}
 2^{2R} \leq   1+\gamma_2\gamma_4\beta_{3t}^2{P^2} +  2\gamma_2\gamma_4\beta_{3t}{P^2} (\beta_{3}-\beta_{3t})  +\beta_3P(\gamma_2+\gamma_4).
\end{equation}
Therefore, the optimization problem is further reformulated as
\begin{subequations}
\begin{align}
&\max \limits_{\beta_1, \beta_2, \beta_3, R }~~ R  \\
&~~~~\text{s.t.}~~~~~~~ 0< \beta_1, \beta_2, \beta_3 < 1,~~\beta_1+\beta_2+\beta_3=1,  \\
&~~~~~~~~~~~~~~  R \leq R_{MAC},~~(\ref{18d}),~~(\ref{18e}),~~(\ref{18f}),
\end{align}
\end{subequations}
which is a convex optimization problem and can be solved efficiently via CVX.

\subsection{Proposed Max-Min-SR}
In the subsection III-A, Max-SR is proposed to enhance the rate performance. Meanwhile, Max-SR needs operations to convert constraints from non-convex to convex. To further improve the sum rate gain, Max-Min-SR method is proposed, where two intermediate variables $R_1$ and $R_2$ are introduced. The optimization problem can be given by
\begin{subequations}
\begin{align}
&\max \limits_{\beta_1, \beta_2, \beta_3 }R=\text{min} \{ \text{min}\{R_{1ir},R_{ri2}  \}+\text{min}\{R_{2ir},R_{ri1}  \} \nonumber\\
&~~~~~~~~~~~~~~~~~~~~~~~~~~~~~~~~~~~~~~~~~~~~~~~~~~~~      , R_{MAC}  \}  \\
&~~~\text{s.t.}~~~ 0< \beta_1, \beta_2, \beta_3 < 1,~~\beta_1+\beta_2+\beta_3=1.
\end{align}
\end{subequations}
It is defined that $  R_{1ir}\geq R_1 $, $  R_{ri2} \geq R_1 $, $  R_{2ir}\geq R_2 $, $  R_{ri1} \geq R_2 $, $  R_{MAC} \geq R $ and $ R_1+R_2 \geq R$. The above optimization problem is reformulated as
\begin{subequations}
\begin{align}
&\max \limits_{ \beta_1,\beta_2,\beta_3,R_1,R_2,R}R \\
&~~~~~~~\text{s.t.}~ 0< \beta_1, \beta_2, \beta_3 < 1,~~\beta_1+\beta_2+\beta_3=1, \\
&~~~~~~~~~~~ 2^{2{R_1}} \leq   1+\gamma_1\beta_1P  , ~~ 2^{2{R_1}} \leq   1+\gamma_2\beta_3P ,\\
&~~~~~~~~~~~ 2^{2{R_2}} \leq   1+\gamma_3\beta_2P  , ~~ 2^{2{R_2}} \leq   1+\gamma_4\beta_3P ,\\
&~~~~~~~~~~~ 2^{2R} \leq   1 + \gamma_1\beta_1P + \gamma_3\beta_2P  , ~  R \leq R_1+R_2,
\end{align}
\end{subequations}
where the object function and each constraint are convex. Thus it is a convex optimization problem, which can also be solved by CVX directly.

\subsection{Proposed Max-SR-RC}
In fact, there exist asymmetry of two-way channel quality and two users' demand in the IRS-aided two-way DF relay network, thereby the asymmetry between $R_{12}$ and $R_{21}$ is generated. In this subsection, we make an investigation of PA in the case of $R_{12}=\mu R_{21}$, which is called Max-SR-RC method. The optimization problem is given by
\begin{subequations}
\begin{align}\label{C}
&\max \limits_{\beta_1, \beta_2, \beta_3 } R  \\
&~~\text{s.t.}~~~~~ 0< \beta_1, \beta_2, \beta_3 < 1,~~\beta_1+\beta_2+\beta_3=1,   \label{C_1}\\
&~~~~~~~~~~  0<\mu,~~R_{12}={\mu}R_{21}, \label{C_2}
\end{align}
\end{subequations}
where $\mu$ is a constant. Substituting (\ref{C_2}) into the object function and expanding it, the above problem can be rewritten as
\begin{subequations}
\begin{align}\label{D}
&\max \limits_{\beta_1, \beta_2, \beta_3 } \text{min} \{ \frac{1}{2}\log_2 (1+ \gamma_3\beta_2P)^{1+\mu} , \frac{1}{2}\log_2 (1+ \gamma_4\beta_3P)^{1+\mu}  \nonumber\\
&~~~~~~~~~~~~~~~~~~~~~~~~~~, \frac{1}{2}\log_2 (1+ \gamma_1\beta_1P + \gamma_3\beta_2P) \} \\
&~~~\text{s.t.}~~~~ 0<\mu,~~(\text{\ref{C_1}}).\label{D_1}
\end{align}
\end{subequations}
Since it is similar to Max-SR method, the above problem can be further transformed as
\begin{subequations}
\begin{align}\label{E}
&\max \limits_{ \beta_1,\beta_2,\beta_3,R}~~  R \\
&~~~~\text{s.t.}~~~~~~~   2^{2R}\leq  (1+\gamma_3\beta_2P)^{1+\mu} , \label{E_1}\\
&~~~~~~~~~~~~~~ 2^{2R}\leq  (1+\gamma_4\beta_3P)^{1+\mu} , \label{E_2}\\
&~~~~~~~~~~~~~~ 2^{2R}\leq   1 + \gamma_1\beta_1P + \gamma_3\beta_2P ,~~(\text{\ref{D_1}}),
\end{align}
\end{subequations}
where (\ref{E_1}) and (\ref{E_2}) are non-convex constraints. In the same manner, the low bounds of convex functions $(1+\gamma_3\beta_2P)^{1+\mu}$ and  $(1+\gamma_4\beta_3P)^{1+\mu}$ can be achieved through the first-order Taylor expansion. The details are as follow
\begin{align}
&(1+\gamma_3\beta_2P)^{1+\mu} \geq  (1+\gamma_3{\beta_{2t}}P)^{1+\mu}  \nonumber\\
&~~~~~~~~~~~~~~~~~~~~~ + (1+\mu)(1+\gamma_3{\beta_{2t}}P)^{\mu}( \beta_2-{\beta_{2t}} ), \\
&(1+\gamma_4\beta_3P)^{1+\mu} \geq  (1+\gamma_4{\beta_{3t}}P)^{1+\mu}  \nonumber\\
&~~~~~~~~~~~~~~~~~~~~~ + (1+\mu)(1+\gamma_4{\beta_{3t}}P)^{\mu}( \beta_3-{\beta_{3t}} ),
\end{align}
where $\beta_{2t}$ and $\beta_{3t}$ are feasible points. Substituting the two low bounds into the above optimization problem, respectively, yields
\begin{subequations}
\begin{align}
&\max \limits_{ \beta_1,\beta_2,\beta_3,R}R \\
&~~~~\text{s.t.}~~~~~  2^{2R}\leq  (1+\gamma_3{\beta_{2t}}P)^{1+\mu}  \nonumber\\
&~~~~~~~~~~~~~~~~~~~~ + (1+\mu)(1+\gamma_3{\beta_{2t}}P)^{\mu}( \beta_2-{\beta_{2t}} ), \\
&~~~~~~~~~~~~~ 2^{2R}\leq  (1+\gamma_4{\beta_{3t}}P)^{1+\mu}  \nonumber\\
&~~~~~~~~~~~~~~~~~~~~ + (1+\mu)(1+\gamma_4{\beta_{3t}}P)^{\mu}( \beta_3-{\beta_{3t}} ), \\
&~~~~~~~~~~~~~ 2^{2R}\leq   1 + \gamma_1\beta_1P + \gamma_3\beta_2P ,~~(\text{\ref{D_1}}),
\end{align}
\end{subequations}
which is also a convex optimization problem. Similarly, PA factors $\beta_1$, $\beta_2$, $\beta_3$ and sum rate $R$ can be obtained.

\section{Simulation And Numerical Results}
In this section, numerical simulations are performed to evaluate and compare the sum rate performance between the proposed three methods and EPA method.
Moreover, it is assumed that $U_1$, $U_2$, IRS and RS are located in three-dimensional (3D) space, the positions of $U_1$, $U_2$, IRS and RS are given as (0, 0, 0), (0, 100m, 0), ($-$10m, 50m, 20m) and (10m, 50m, 10m), respectively. We take a more realistic environment into account, and assume that all channels follow large-scale fading, which contains shadow fading. The path loss model is $PL(d)=PL_0-10{\alpha}\text{log}_{10}(\frac{d}{d_0})-X_\sigma$, where $PL_0=-20\text{log}_{10}(\frac{4{\pi}d_0}{\lambda})$ is the path loss at the reference distance $d_0$, $\lambda$ is wavelength, $\alpha$ is the path loss exponent, and $X_\sigma$ is a Gaussian random shadow variable with distribution $X_\sigma\sim\mathcal{CN}( 0,\sigma^2 )$. The path loss exponents associated with IRS are 2.1, those of $U_1$-RS and $U_2$-RS links are 2.3. The remaining system parameters are set as follows: $f_c=$ 1.5GHz, $\sigma_{1}^2=\sigma_{2}^2=\sigma_{r}^2= -$80dBm.

\begin{figure}[h]
\centering
\includegraphics[width=0.45\textwidth,height=0.2673\textheight]{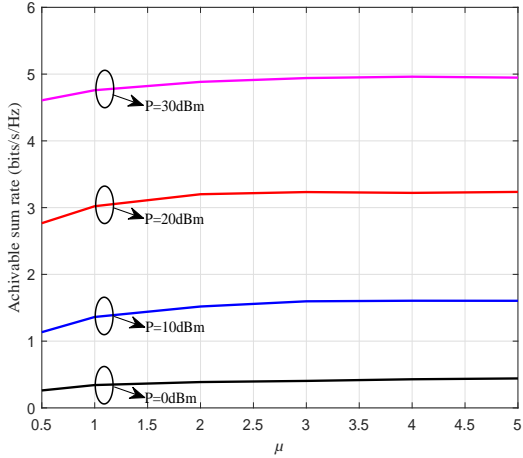}\\
\caption{Achievable sum rate of proposed Max-SR-RC versus $\mu$ with $M=$ 4, $N=$ 16 and $\sigma=3$dB.}\label{Rate_VS_u.eps}
\end{figure}

\begin{figure}[h]
\centering
\includegraphics[width=0.45\textwidth,height=0.2673\textheight]{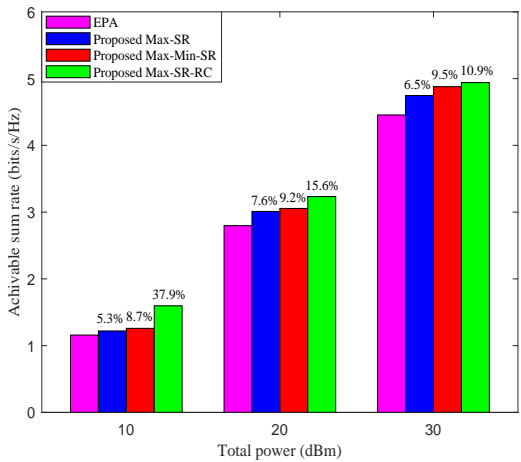}\\
\caption{Achievable sum rate versus total power with $M=$ 4, $N=$ 16 and $\sigma=3$dB.}\label{bar_Rate_VS_Power.eps}
\end{figure}

\begin{figure}[h]
\centering
\includegraphics[width=0.45\textwidth,height=0.2673\textheight]{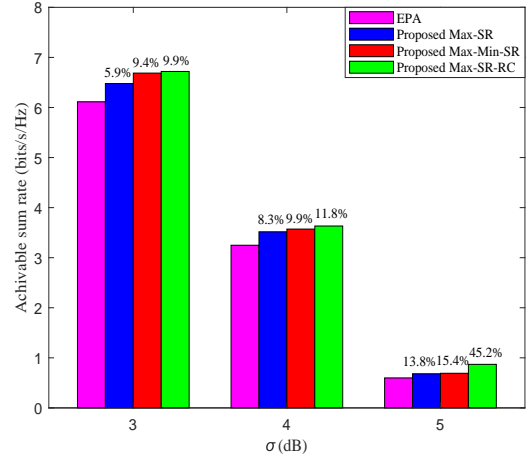}\\
\caption{Achievable sum rate versus $\sigma$ with total power $=$ 40dBm, $M=$ 4 and $N=$ 16.}\label{bar_Rate_VS_sigma.eps}
\end{figure}

Fig.~2 illustrates the achievable sum rate of Max-SR-RC method versus $\mu$ with different total power: $P = \{$0dBm, 10\text{dBm}, 20\text{dBm}, 30\text{dBm}$\}$. It can be seen that as $\mu$ increases, the sum rate performance gradually increases. Until $\mu \geq$ 3, the achievable sum rate tends to be stable. For convenience, $\mu$ is defined as 3.

Fig.~3 depicts the histogram of achievable sum rate versus total power with $M=$ 4, $N=$ 16 and $\sigma=$ 3dB. It can be seen that the proposed three methods perform better than EPA method. For instance, when  total power equals 10dBm, the proposed worst method, Max-SR, can harvest up to 5.3\% rate gain over EPA method. The best method, Max-SR-RC, approximately has a 37.9\% rate gain over EPA method.

Fig.~4 shows the corresponding histogram of achievable sum rate versus $\sigma$ with total power $=$ 40dBm, $M=$ 4 and $N=$ 16. When $\sigma=$ 5dB, it can be observed that the achievable sum rate performance improvements over EPA method are 13.8\%, 15.4\% and 45.2\%, respectively. Moreover, it indicates that as $\sigma$ increases, the sum rate performance gain becomes more obvious, which is very attractive. However, greater shadowing fading is formed because of the increase of $\sigma$, which sharply deteriorates the sum rate performance.

\section{Conclusions}
In this paper, in order to improve the achievable sum rate of an IRS-aided two-way DF relay wireless network, three high-performance PA schemes, namely Max-SR, Max-Min-SR and Max-SR-RC, were proposed. Simulation results showed that the proposed three methods can harvest better rate gain over EPA method in terms of sum rate performance. Because of excellent rate performance, the proposed Max-SR-RC method is very attractive. Additionally, the rate increases with total power, while decreases with $\sigma$.

\ifCLASSOPTIONcaptionsoff
  \newpage
\fi

\bibliographystyle{IEEEtran}
\bibliography{IEEEfull,reference}
\end{document}